\documentclass[aps,prd,superscriptaddress,10pt,showpacs,notitlepage, twocolumn, 
tightnlines
]{revtex4-1}
	\usepackage{graphicx}
	\usepackage{amsmath,amssymb,mathrsfs}
\usepackage{epsf}
\usepackage{epsfig}
\usepackage{mathptmx}

\usepackage{color}

\usepackage{amsmath}

\usepackage{amssymb}
\usepackage{bm}
\numberwithin{equation}{section}

\usepackage{subfigure}

\newcommand{\p}{\partial}

\def\L{\left}
\def\R{\right}

\def\half{\frac{1}{2}}
\DeclareMathAlphabet{\mathpzc}{OT1}{pzc}{m}{it}
\def\r{\tilde{\rho}}

\def\t{\theta}
\def\d{\delta}

\graphicspath{{Fig/}}
\def\g{\bm{g}}

\DeclareMathAlphabet{\mathpzc}{OT1}{pzc}{m}{it}

\begin{document}
\title{Collective excitations of a quantized vortex in $^3P_2$ superfluids in neutron stars}

\date{\today}

\author{Chandrasekhar Chatterjee}
\email{chandra@phys-h.keio.ac.jp}

\affiliation{%
Department of Physics, and Research and Education Center for Natural Sciences, Keio University, Hiyoshi 4-1-1, Yokohama, Kanagawa 223-8521, Japan
}%

\author{Mareike Haberichter}%
 \email{mareike@math.umass.edu}

\affiliation{%
Department of Physics, and Research and Education Center for Natural Sciences, Keio University, Hiyoshi 4-1-1, Yokohama, Kanagawa 223-8521, Japan
}%

\affiliation{%
Institut f{\"{u}}r Physik, Universit{\"{a}}t Oldenburg, Postfach 2503 D-26111 Oldenburg, Germany
}%

\affiliation{%
Department of Mathematics and Statistics, University of Massachusetts, Amherst, Massachusetts 01003-4515, USA
}%

\author{Muneto Nitta}%
\email{nitta@phys-h.keio.ac.jp}

\affiliation{%
Department of Physics, and Research and Education Center for Natural Sciences, Keio University, Hiyoshi 4-1-1, Yokohama, Kanagawa 223-8521, Japan
}%

\begin{abstract}
We discuss collective excitations (both fundamental and solitonic excitations)
of quantized superfluid vortices in neutron $^3P_2$ superfluids, which likely exist in high density neutron matter such as neutron stars. Besides the well-known Kelvin modes (translational zero modes), we find a gapfull mode whose low-energy description takes the simple form of a double sine-Gordon model. The associated kink solution and its effects on spontaneous magnetization inside 
the vortex core are analyzed in detail.
\end{abstract}

\maketitle

\section{Introduction}
The properties of high density nuclear matter are still not sufficiently understood. However, in recent years, neutron star (NS) observations began to place stronger restrictions  \cite{Steiner:2012xt,Lattimer2007109,lrr-2008-10} on the equation of state (EoS) of dense matter and to probe structure and composition of NS cores. Most valuable information into the state of their interiors came from the detections of pulsar glitches via pulsar timing measurements, optical and X-ray observations of cooling and accreting NSs and from neutrino emission measurements from proto-neutron stars. These observations provide evidence for superfluidity in the interiors of NSs \cite{saullecture}. 

The first direct evidence \cite{PhysRevLett.106.081101,Shternin01032011} that the NS core should exist in a superfluid state has been reported in the study of the NS in the supernova remnant known as Cassiopeia A. It has been measured that Cassiopeia A's surface temperature has rapidly decreased from $2.12\times 10^6$ K to $2.04\times 10^6$ K. This pronounced drop in surface temperature can be naturally explained \cite{PhysRevLett.106.081101,Shternin01032011,Leinson:2014cja} if one assumes that neutrons have recently become superfluid in the NS core.  As the neutrons combine to form Cooper pairs, 
a splash of neutrinos is emitted accelerating the NS cooling process. Another evidence for a superfluid NS core comes from the observation of pulsar glitches, which are sudden jumps in the NS rotation frequency. Glitches like the ones observed in the Crab and Velar pulsar can be explained by two main physical mechanism -- either they are due to starquakes from the NS core \cite{Pines:1972,Takatsuka01011988} or crust  \cite{Baym:1969,Anderson:1975zze,1976ApJ...203..213R} or they are caused by the sudden unpinning and displacement of a large number of vortices \cite{Anderson:1975zze,1976ApJ...203..213R,Pines1985} in the NS superfluid. In the starquake glitch model \cite{1976ApJ...203..213R,Baym:1971,1996ApJ...459..706A}, the glitches are caused by a sudden reduction in the moment of inertia of some solid component of the NS. For example, the differential rotation between the crust and the superfluid NS core can produce stresses \cite{1976ApJ...203..213R} in the NS crust leading to crustquakes. The resulting crustquakes distort the star's shape and hence generate sudden jumps in the NS rotational frequency, seen as glitches. In contrast, in the pinned superfluid model angular momentum is suddenly transferred from the superfluid NS core to the non-superfluid crust via vortex unpinning, spinning it up. Neutron superfluid vortices can become pinned to nuclear clusters in the inner crust \cite{Anderson:1975zze} or to magnetic flux tubes in the core \cite{Baym:1969nat}. This prevents angular momentum from being transferred to the NS crust. Hence, a differential rotation builds up between the NS core and crust. When this differential becomes large enough angular momentum is suddenly transferred from the core to the crust through the catastrophic unpinning of vortices, resulting in a sudden spin-up in the NS. Furthermore, the observed long time relaxation after glitches
can be explained only by assuming the coexistence of normal and superfluid components \cite{Baym:1969,Pines:1972}. Therefore, understanding the properties of neutron superfluidity and the formation and dynamics of superfluid vortices can give us further insights into the evolution of NSs and their composition.

At the high density central part of NSs the neutron superfluidity is attributed to the $^3 P_2$ pairing interaction \cite{Tamagaki01101970,PhysRevLett.24.775,Takatsuka01071971,Takatsuka01031972,Fujita01091972,Richardson:1972xn} rather than to $^1 S_0$ pairing familiar from the conventional BCS theory of superconductors.  Analysis of nucleon-nucleon scattering data  \cite{PhysRevLett.24.775} shows that the transition from an isotropic  $^1 S_0$ superfluid to an anisotropic $^3P_2$ superfluid occurs at densities above $1.6\times 10^{14}\, g/\text{cm}^3$. The Ginzburg-Landau (GL) energy functional generalized for $^3P_2$ superfluids was developed in Refs.~\cite{Fujita01091972,Richardson:1972xn} in the weak coupling limit. 
Note that the general GL free energy formally agrees with the angular momentum $l=2$ GL functional solved by Mermin \cite{PhysRevA.9.868};
depending on the GL parameters, 
the ground state is in either nematic, ferromagnetic or cyclic phase. 
The ground states of $^3P_2$ superfluids in the weak coupling limit 
were found in Ref.~\cite{PhysRevD.17.1524} 
to be in the nematic phase in the absence of a magnetic field. 
They are continuously degenerated when we ignore the sixth order term 
and can be decomposed by unbroken 
$O(2)$, $D_2$ and $D_4$ symmetry groups 
into the uniaxial, $D_2$- and $D_4$-biaxial nematic phases, respectively.   
(This is also known from spin-2 spinor Bose-Einstein condensates, see Refs.~\cite{Song:2007ca,Uchino:2010pf}.) 
This  degeneracy is lifted and the uniaxial nematic phase
becomes the unique ground state once the sixth order term is taken into account. 
The ground states in the presence of a magnetic field 
were recently determined \cite{Masuda:2015jka};
the ground state is in the uniaxial nematic phase for 
smaller magnetic fields relevant for ordinary NSs 
and in the $D_2$- or $D_4$-biaxial nematic phase  
for large magnetic field relevant for magnetars. 
Beyond the GL theory, 
the ground states in the $(T,H)$ phase diagram 
have been obtained recently
in the Bogoliubov-de Gennes (BdG) formalism 
\cite{Mizushima:2016fbn}. 
The $D_2$- and $D_4$-biaxial nematic phases
appear in lower $T$ and $H$ region and 
in higher $T$ and $H$ region, respectively.
The phase boundary is of second order at higher temperature 
while it is of first order at lower temperature, 
and a tricritical point connects these boundaries.
One of the most important results of the BdG formalism is that 
$^3P_2$ superfluids were found to be topological superfluids 
predicting gapless Majorana fermions on the surface 
\cite{Mizushima:2016fbn}. The strong-coupling corrections to the $^3 P_2$ NS matter GL free energy including spin-orbit and central forces were calculated in Ref.~\cite{PhysRevD.29.2705}.

As phenomenological aspects relevant for NSs physics are concerned, 
$^3P_2$ superfluids provide new mechanisms for neutrino emission of dense neutron matter  \cite{Bedaque:2003wj,Bedaque:2012bs,Bedaque:2014zta,Leinson:2009nu,Leinson:2010yf,Leinson:2010pk,Leinson:2010ru,Leinson:2011jr,Leinson:2012pn,Leinson:2013si}, explain the entrainment \cite{Shahabasyan2011} of superconducting protons by rotating superfluid neutrons in NSs and offer possible explanations of the anomalously rapid cooling of NSs \cite{PhysRevLett.106.081101,Shternin01032011,Leinson:2014cja}.

Since superfluids are rotating inside NSs, a large number ($\sim 10^{19}$) of superfluid vortices exist in their interior.  The rich structure and magnetic properties of vortices emerging in the GL equations for  $^3 P_2$ superfluids were explored in Refs.~ \cite{Richardson:1972xn,Muzikar:1980as,Sauls:1982ie,Masuda:2015jka,Masuda:2016vak}. 
$^3P_2$ vortices in NS matter turn out to be structurally different from their counterparts in $^1 S_0$ superfluids. For example, different to $^1 S_0$ vortices, $^3 P_2$ vortices exhibit spontaneous magnetization in the vortex core region
\cite{Sauls:1982ie,Masuda:2015jka}. 
For magnetic field strengths of orders of magnitude as they appear in magnetars, the ground state 
is in the $D_4$-biaxial nematic phase, in which 
the first homotopy group $\pi_1$ is non-Abelian, 
thereby admitting non-Abelian (non-commutative) vortices 
which carry half-quantized circulation 
\cite{Masuda:2016vak}.

In this article, 
we study low-energy collective modes [or (pseudo) moduli]  and solitonic excitations of
an integer vortex in a $^3P_2$ superfluid.
By solving numerically the GL equation, 
we reconstruct the axially symmetric integer vortex solution 
in the absence of magnetic fields and sixth order terms. 
Because of the off-diagonal elements of the tensor order parameter, 
there exists spontaneous magnetization at the vortex core 
as described above. 
Here, we calculate the net magnetic moment per  femtometer along vortex line to be one order less than the neutron magnetic moment.
We then study collective modes in the presence of a 
single vortex. 
As usual, there are gapless (massless) Kelvin modes 
(translational moduli) due to the spontaneously broken 
translational symmetry in the presence of the vortex.
In addition, the phase $\delta$ of the off-diagonal elements of the 
tensor order parameter gives rise to a gapfull (massive) mode. 
We construct the low-energy (long distance) effective free energy 
of this gapfull mode and find it is a double sine-Gordon model, 
consisting of the potential terms cos $\delta$ and cos $2\delta$. 
For GL parameter values describing typical NSs,
it allows only $2\pi$ kink solutions \footnote{
Sine-Gordon kinks on vortices were studied in 
the Skyrme model \cite{Gudnason:2014hsa,Gudnason:2016yix}. 
In this case, 
the kink represents a Skyrmion, carrying a non-zero baryon number.
}. 
We find that the core magnetization of the vortex flips its direction 
at the kink.

The article is structured as follows.
After introducing the GL free energy relevant for $^3P_2$ superfluids in the weak coupling limit, 
$^3P_2$ vortex solutions are constructed in Section~\ref{sec:GL1} and their spontaneous magnetization in the vortex core region is evaluated. Then, in Section~\ref{sec:EffAct}, we discuss the effect of the parameter $\delta$ on the free energy density by writing down the associated effective free energy functional which takes the simple form of a double sine-Gordon model. The associated kink soliton solution is derived in Section~\ref{sec:DGkink}.  Finally, our conclusions and possible future lines of investigation are presented in Section~\ref{sec:Con}. To facilitate the reader to reproduce our numerical results, we add two appendices to this article. In Appendix~\ref{sec:GL}, we briefly review the GL theory for $^3P_2$ superfluid states and state all the parameter values used in our numerical simulations. We list explicitly all the vortex equations together with the imposed boundary conditions in Appendix~\ref{App_Eom}. Note that a detailed investigation of vortex structure and dynamics in the presence of  nonzero external magnetic fields and higher order terms, in particular the inclusion of the sixth order term, will be published in a forthcoming paper.

\section{\label{sec:GL1}Ginzburg-Landau Description of Vortices in   $^3P_2$ Neutron Superfluids}

In this section we discuss the GL free energy and determine the vortex configurations numerically. A discussion of spontaneous magnetization is included at the end of the section.

\subsection{Ginzburg-Landau Free Energy}

The GL free energy for the $^3P_2$ superfluidity was originally derived in 
Refs.~\cite{Fujita01011973, Richardson:1972xn, Sauls:1982ie, Masuda:2016vak} by generalizing Gor'kov's procedure.  In this case the original $^1S_0$ contact interaction was generalised to a $^3P_2$ contact interaction by introducing a derivative coupling. A short explanation of the derivation is given in Appendix~\ref{sec:GL}.

The tensorial order parameter $A_{\alpha i}$ transforms under the symmetry group as
\begin{align}
 A \to e^{i \theta} \g A \g^T, \quad \text{ with }\quad e^{i \theta} \in U(1),
\quad \text{ and }\quad \g \in SO(3)_{L+S}.
\end{align}
Here, $SO(3)_{L+S}$ is the diagonal subgroup of the full group $SO(3)_{L} \times SO(3)_{S}$ and  is generated by the total angular momentum $J = L + S$.

The GL free energy density $F$  can be written as a function of the tensor $A_{\mu i}$ 
\begin{eqnarray}\label{Free}
F= \frac{1}{G}\int d^3 \rho \ \L(F_{\rm grad} + F_{2+4}+F_6+F_H\R), \label{freeenergydensity}
\end{eqnarray}
where $F_{\rm grad}$ is the gradient term, 
$F_{2+4}$ and $F_6$  \cite{saulsthesis:1980ab} are the free energy densities up to fourth order and up to sixth order, 
respectively and $F_H$ is the magnetic term. Here, we rescaled for later convenience the free energy functional by $G=3\pi^2/\left(m_n k_F^3\right)\approx  115.39\,\text{MeV}\,\text{fm}^5$ where $k_F$ is the Fermi momentum and $m_n$ represents the  nucleon mass, see Appendix~\ref{sec:GL} for the numerical values of all model parameters.  
The free energy density contributions are explicitly given by 
\begin{subequations}\label{free_contri}
\begin{align}
 F_{\rm grad} &= K_1\, \partial_iA_{\alpha j}\partial_iA^{\dagger}_{\alpha j} 
 + K_2\, \L(\partial_iA_{\alpha i}\partial_jA^{\dagger}_{\alpha j}+\partial_iA_{\alpha j}\partial_jA^{\dagger}_{\alpha i}\R) , \label{free_contri_grad}\\
F_{2+4} &= \alpha\,\, {\rm Tr}AA^{\dagger}
 +\beta\,\,\L[\L({\rm Tr}AA^{\dagger}\R)^2-{\rm Tr}A^2A^{\dagger 2}\R], \label{4th_contri_pot}  \\
F_6 =&
\gamma_6 \,\L[-3\L({\rm Tr}AA^{\dagger}\R)|{\rm Tr}AA|^2
+4({\rm Tr}AA^{\dagger})^3  
\R. \nonumber\\&\L. 
+12({\rm Tr}AA^{\dagger}){\rm Tr}(AA^{\dagger})^2 
+ 6({\rm Tr}AA^{\dagger}){\rm Tr}(A^2A^{\dagger 2})\R.\qquad \nonumber\\&\L. 
+8{\rm Tr}(AA^{\dagger})^3 +12{\rm Tr}[(AA^{\dagger})^2A^{\dagger}A]\R. \nonumber \\ 
&\L. 
-12{\rm Tr}[AA^{\dagger}A^{\dagger}A^{\dagger}AA]
-12{\rm Tr}AA({\rm Tr}AA^{\dagger}AA)^{\ast}\R],\\
F_H &= g'_H\, H^2 {\rm Tr}\L(A A^{\dagger}\R)+g_H \,H_{\alpha}\L(AA^{\dagger}\R)_{\alpha \beta}H_{\beta},
\label{eq:fB}
\end{align}
\end{subequations}
where $H$ is the strength of the external magnetic field and we implicitly sum over repeated indices. Note that for simplicity, we set $g'_H=0$ in Eq.~(\ref{eq:fB}). In this paper we also neglect the sixth order term because of small values of $\gamma_6$. Existence of nonzero $\gamma_6$ would make our system  metastable, however we assume that small value of $\gamma_6$ would make the relaxation time much longer than the time scale of the system. All the following calculations will be performed in the weak coupling limit by considering only the excitations around the Fermi surface \cite{Fujita01011973,Richardson:1972xn, saulsthesis:1980ab}. In this limit, $K_1$ and $K_2$ in Eq.~(\ref{free_contri_grad}) take the same value which is set to be $K$ in the following. To simplify comparison with the works \cite{Richardson:1972xn, Sauls:1982ie, Masuda:2015jka,Masuda:2016vak}, we fix the numerical values of the model parameters appearing in Eq.~(\ref{free_contri}) as discussed in Appendix~\ref{sec:GL} and Table~\ref{table-coeffi}.

As mentioned in the introduction, the ground state of the GL free energy is continuously degenerate in the absence of $\gamma_6$ and of external magnetic fields. Namely, the ground state order parameter in the cartesian basis can always be diagonalised using $SO(3)$ and $U(1)$ rotations to the form 
\begin{eqnarray}
\label{gs}
A_{gs} = \sqrt{\frac{|\alpha|}{\beta(1 + \eta^2 + (1 + \eta)^2)}}
\left(
\begin{array}{ccc}
\eta  & 0  & 0  \\
0  &-(1+\eta)   & 0  \\
 0 &  0 & 1  
\end{array}
\right)\,,
\end{eqnarray}
where $\eta$ is a dimensionless numerical parameter \cite{Sauls:1982ie,PhysRevLett.97.180412,Masuda:2015jka} taking values within the interval  $-1 \le \eta \le -\frac{1}{2}$.

\subsection{Vortex Configuration}\label{sec:Vortex}
Vortex configurations in the absence of a magnetic field and sixth order terms 
were first studied in Refs.~\cite{Richardson:1972xn,Muzikar:1980as}.
Since the order parameter is tensorial, we can choose a basis which diagonalizes the tensor at large distances from the vortex core.
In Refs.~\cite{Richardson:1972xn,Muzikar:1980as}, the configuration in which 
the tensor is diagonalized in the cylindrical basis was considered. Later, this configuration 
was compared with the configuration in which the tensor is diagonalized in 
the Cartesian basis, and it was confirmed that the former gives the lowest energy 
configuration 
 \cite{Masuda:2015jka}.

One interesting feature is that the vortex profile functions eventually choose asymptotically only one point among the whole degenerate ground state values. This means that the continuos degeneracy  of the ground state, as mentioned in Eq.~(\ref{gs}),  is lifted \cite{Richardson:1972xn, Sauls:1982ie, Masuda:2015jka} in the presence of a vortex at large distances.

In this subsection, we shall derive the equations of motion for the vortex profile functions using the following ansatz \cite{Masuda:2015jka} for the order parameter of a vortex state 
\begin{eqnarray}
&& A^{(x,y,z)} 
= \sqrt{\frac{|\alpha|}{6\beta}}R(\theta)A^{(\rho, \t, z)}R^T(\theta)e^{i  \theta}, \nonumber \\
&& A^{(\rho, \t, z)} =
  \left(
    \begin{array}{ccc}
      f_1 & ige^{i\delta} & 0 \\
      ige^{ i\delta} & f_2 & 0 \\
      0 & 0 & -f_1-f_2
    \end{array}
  \right),    \label{eq:ansatz_1}
\end{eqnarray}
where $(\rho,\theta,z)$ denote cylindrical coordinates and $\delta$ is a constant parameter. 
$A^{(x,y,z)}$ is the order parameters in the Cartesian basis and 
$A^{(\rho,\t,z)}$ is the order parameters in 
the cylindrical basis
which are related by 
a rotation matrix  $R$, given by 
\begin{eqnarray}
R(\theta)=
\left(
    \begin{array}{ccc}
      {\rm{cos}}\theta & -{\rm{sin}}\theta & 0 \\
      {\rm{sin}}\theta & {\rm{cos}}\theta & 0 \\
      0 & 0 & 1
    \end{array}
  \right) .
\end{eqnarray} 
For a brief motivation of the ansatz (\ref{eq:ansatz_1}), we refer the reader to Appendix.~\ref{vortex:order} and Refs.~\cite{Fujita01091972,Richardson:1972xn,Sauls:1982ie,saulsthesis:1980ab}. In Eq.~(\ref{eq:ansatz_1}), $f_1$, $f_2$ and $g$ are profile functions depending only on the radial coordinate $\rho$, and satisfying the boundary conditions
\begin{align}\label{BCs}
 & f_1\to -\sqrt{\frac{6}{2\eta^2+2\eta+2}}\eta=1.1116\,,\quad \mbox{as } 
 \rho \to \infty\, \nonumber\\
 & f_2 \to  \sqrt{\frac{3}{2}}\frac{2\eta+2}{\sqrt{2\eta^2+2\eta+2}}= 0.8841\quad \mbox{as } 
 \rho \to \infty\, ,\nonumber \\
 & g \to 0  \quad \mbox{as } 
 \rho \to \infty\,  \\
 & f_1,f_2 \to 0, 
 \quad  g \to 0 \,\,\,\,\quad \quad\quad \quad   \quad \quad \quad \mbox{as } \rho \to 0\,,
\end{align}
where the dimensionless parameter $\eta$ has a boundary value at spatial infinity determined by energy minimisation 
to be given by $\eta = 6 -\sqrt{43} \sim -0.557$
\cite{Richardson:1972xn,Muzikar:1980as,Masuda:2015jka};
the continuous degeneracy of the ground state is lifted
in the vortex background as described above. As the profile function $g$ vanishes at the origin and at spatial infinity, it creates ring-shaped soliton solutions and in general one may replace  $g$  by $g e^{im\theta}$, where $m$ denotes how many times the phase of $g$ is twisted along the ring. Note that in Ref.~\cite{Kobayashi:2012ib} winding numbers that are locally defined in the core region were used to classify vortex-core structures in spin-1 Bose-Einstein condensates.

In Refs.~\cite{Masuda:2015jka,Richardson:1972xn,Muzikar:1980as} vortex solutions in $^{3}P_2$ superfluids have been constructed and analysed using the ansatz (\ref{eq:ansatz_1}) for the order parameter of a vortex state. However, the effects of a non-zero phase parameter $\delta$ on the vortex structure have not been considered so far in the existing literature. Here, we introduce the constant phase $\delta$ in (\ref{eq:ansatz_1}) to gain further insight into the collective excitations localised around a vortex. 
Around the vicinity of the vortex, one off-diagonal component was considered so far in the literature \cite{Masuda:2015jka,Richardson:1972xn,Muzikar:1980as}. However, the authors only discussed the implications of a 
purely imaginary off-diagonal element. As a further generalisation of the vortex ansatz, we allow for real off-diagonal element contributions by introducing the non-zero phase $\delta$ in (\ref{eq:ansatz_1}) as a minimal extension. This generalised ansatz is partly motivated by the idea that the phase might be a Nambu-Goldstone mode as it is often the case for solitons that a phase of a localised profile can be identified with a Nambu-Goldstone mode localised around the soliton. However, as we will learn in the following there exists a mass gap and therefore the phase $\delta$ cannot be considered as a Nambu-Goldstone mode.

Using the ansatz (\ref{eq:ansatz_1}) we may  rewrite the  free energy (\ref{Free}) as
\begin{align}\label{Free_en}
F&=\int\text{d}^2\rho\,\frac{|\alpha|}{6\beta}\frac{K}{G}\left\{ \left(t_1+ t_2\right)+ \frac{\alpha}{K} t_3+\frac{|\alpha|}{6 K}t_4\right\}\,,
\end{align}
where the individual energy density contributions $t_1$, $t_2$, $t_3$ and $t_4$ take the following form
\begin{subequations}
\label{t_general_Exa1}
\begin{align}
t_1&=2\left(f_1^{\prime2}+f_2^{\prime2}+f_1^\prime f_2^\prime+g^{\prime2}\right)+\frac{1}{\rho^2}\left(4f_1^2+4f_2^2-2f_1 f_2+10g^2\R.\nonumber \\ &\L.-8\left(f_1-f_2\right)g\cos\delta\right)\,,\\
t_2&=2\left(f_1^{\prime2}+g^{\prime2}\right)+\frac{2}{\rho^2}\left\{f_2^2+5g^2+\left(f_1-f_2\right)^2\R.\nonumber\\&\L.-2g\left(f_1-f_2\right)\cos\delta+4f_2g\cos\delta\right\}
+\frac{1}{\rho}\left\{-2\left(f_1^\prime+f_2^\prime\right)g\cos\delta\R.\nonumber\\&\L.+2\left(f_1+f_2\right)g^\prime\cos\delta+2\left(f_1^\prime+f_2^\prime\right)\left(f_1-f_2\right)\right\}\nonumber,\\
t_3&=2\left(f_1^2+f_2^2+f_1f_2+g^2\right)\,,\\
t_4&=2f_1^4+4f_1^3f_2+6f_1^2f_2^2+4f_1f_2^3+2f_2^4+\Big\{\left(6+2\cos2\delta\right)f_1^2 \nonumber\\&+4f_1f_2+\left(6+2\cos2\delta\right) f_2^2+4 f_1f_2\Big\}g^2+2g^4\,.
\end{align}
\end{subequations}
Here the prime denotes the derivative with respect to $\rho$. For numerical calculations it is useful to rescale the radial coordinate by $\rho \rightarrow \tilde{\rho} = \rho/\xi$ where  $1/\xi^2 = 12 |\alpha|/K$ with $\xi$ being the coherence length ($\sim30$ fm) and to change variables to $u = f_1 +f_2$ and $v=f_1 - f_2$.
After rescaling and change of variables, the free energy (\ref{Free_en}) can be written as
\begin{align}
\label{Free_case_1}
 F &= \int \,\text{d}^2\r\,\frac{2|\alpha|^2 \xi^2}{G \beta}\left\{ \left(\tilde{t}_1+ \tilde{t}_2\right)+ \frac{\alpha}{12|\alpha|} t_3+\frac{1}{72}t_4\right\}
\end{align}
where the energy contributions read
\begin{subequations}
 \begin{align}\label{t_Exa1_uv}
\tilde{t}_1&=\frac{3}{2}u^{\prime2}+\frac{1}{2}v^{\prime2}+2g^{\prime2}+\frac{1}{\r^2}\left(\frac{3}{2}u^2+\frac{5}{2}v^2+10g^2-8vg\cos\delta\right)\,,\\
\tilde{t}_2&=\frac{1}{2}\left(u^\prime+v^\prime\right)^2+2g^{\prime 2}+\frac{1}{\r^2}\left(\frac{1}{2}u^2+\frac{5}{2}v^2-uv+10g^2 \R.\nonumber\\&\L.+ 4g(u - 2v)\cos\delta\right)+\frac{1}{\r}\left( 2(u g' - gu')\cos\delta+2u^\prime v\right)\,,\\
t_3&=\frac{3}{2}u^2+\frac{1}{2}v^2+2g^2\,,\\
t_4&=\frac{1}{8}\left(3u^2+v^2\right)^2+\left[(4u^2 + 2 v^2) + (u^2 + v^2)\cos2\delta\right]g^2+2g^4\,.
\end{align}
\end{subequations}
Here, the prime denotes the derivative with respect to $\tilde{\rho}$. Note that $ \tilde{t}_1 = \xi^2 t_1(\rho) = t_1\left(\rho/\xi\right) $. In the following, we calculate the static vortex solution for $\delta=0$. In the next section, the resulting vortex profile functions will be used to study perturbatively the effects of nonzero $\delta$. We summarise the equations of motion derived from the free energy functional (\ref{Free_case_1}) for vanishing $\delta$ together with the imposed boundary conditions in Appendix~\ref{App_Eom}.  

\begin{figure}[!htb]
\includegraphics[totalheight=7.1cm]{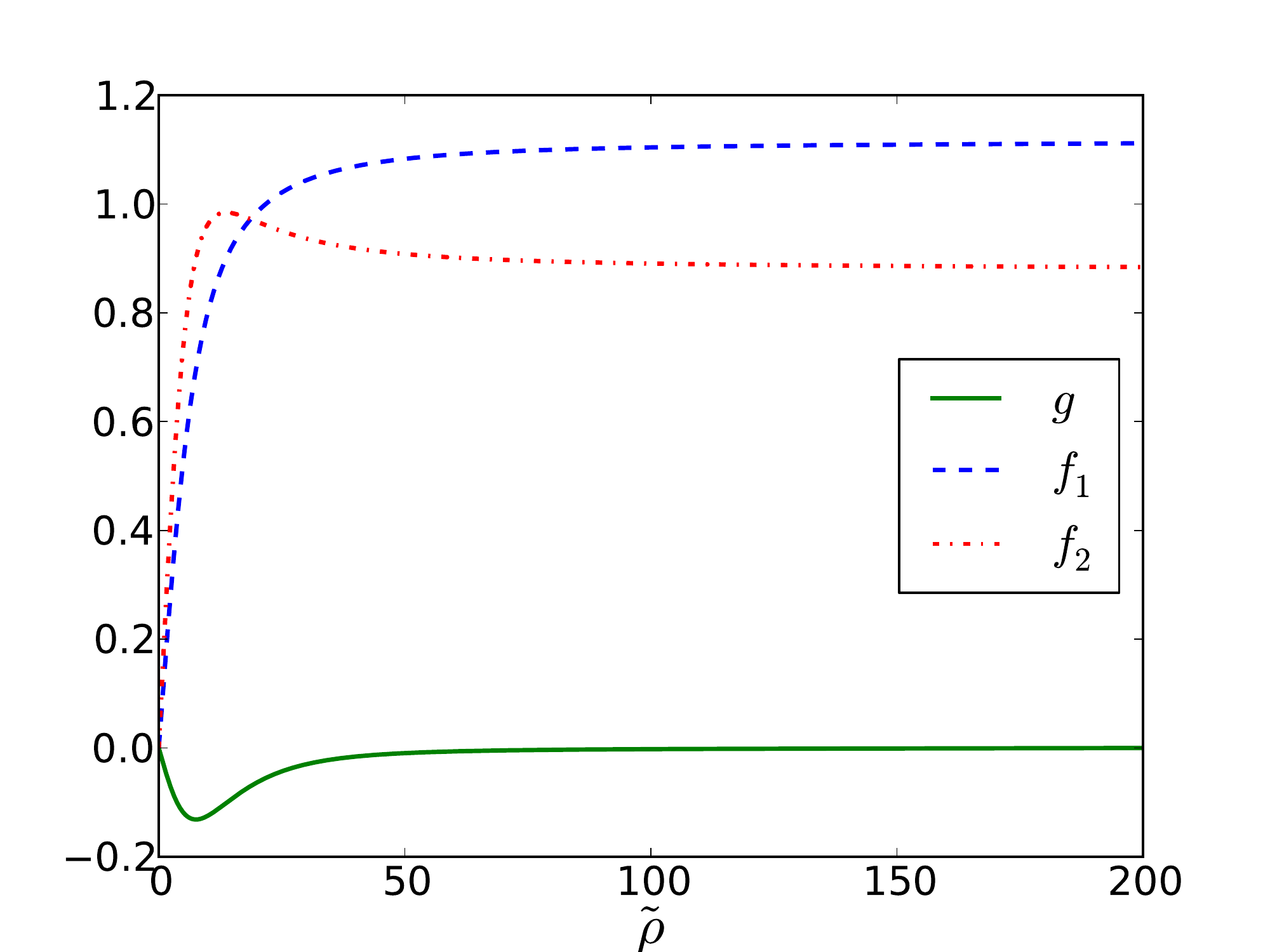}
\caption{We display the profile functions $f_1$, $f_2$ and $g$ for a vortex configuration of winding number one as a function of the distance $\tilde{\rho}$ from the vortex center. Here, the order parameter is expressed in the cylindrical basis ($n=1$) and the vortex solution of lowest energy is found for the boundary value $\eta=-0.557$. Numerical results are shown for the case $m=0$.}
\label{Case1_f1f2_F24}
\end{figure}
The obtained profile functions $f_1$, $f_2$ and $g$ are displayed in Fig.~\ref{Case1_f1f2_F24} as a function of the rescaled radial distance $\tilde{\rho}$ from the vortex center. To obtain solutions such as those shown in Fig.~\ref{Case1_f1f2_F24}, we solve the equations of motion (\ref{EOM_n1m0}) with the boundary conditions stated in Eqs.~(\ref{BC_n1m0}) using a numerical relaxation method. The minimal energy solutions in the model (\ref{Free_case_1}) are constructed by solving the gradient flow equations with a crude initial guess (an approximation in terms of the hyperbolic tangents for the functions $u$ and $v$ and a gaussian curve approximation for the profile function $g$). The initial configurations are relaxed on a spatial grid with 2001 points and spacing $\Delta \tilde{\rho}= 0.1$. 


 $^3 P_2$ vortices are known to differ from their counterparts in $^{1} S_0$ superfluids by their nonzero magnetic moments \cite{Sauls:1982ie,Masuda:2015jka}. From the vortex profile functions displayed in Fig.~\ref{Case1_f1f2_F24}, we can compute the resulting small spontaneous magnetization in the $^3 P_2$ vortex core region. 

The vortex magnetization is given in Refs.~\cite{Sauls:1982ie, Masuda:2015jka} by
\begin{eqnarray}
\bm{M}&=&\frac{\gamma_n \hbar}{2} \bm{ \sigma}\, , 
\label{magnetization}
\end{eqnarray}
where $\bm{\sigma}$ is computed as
\begin{eqnarray}\label{Mag_sig}
\bm{\sigma}&=& 
C_m\frac{|\alpha|}{6\beta}g(\rho)\left[f_1(\rho)-f_2(\rho)\right]{\rm cos}\delta\,\, \hat{\bm{z}}\,.
\end{eqnarray}

In the following, we take $\gamma_n = -1.832 471 72 \times 10^4 s^{-1}G^{-1}$ as experimental value for the neutron gyromagnetic ratio.

\begin{figure}[!htb]
\includegraphics[totalheight=7.1cm]{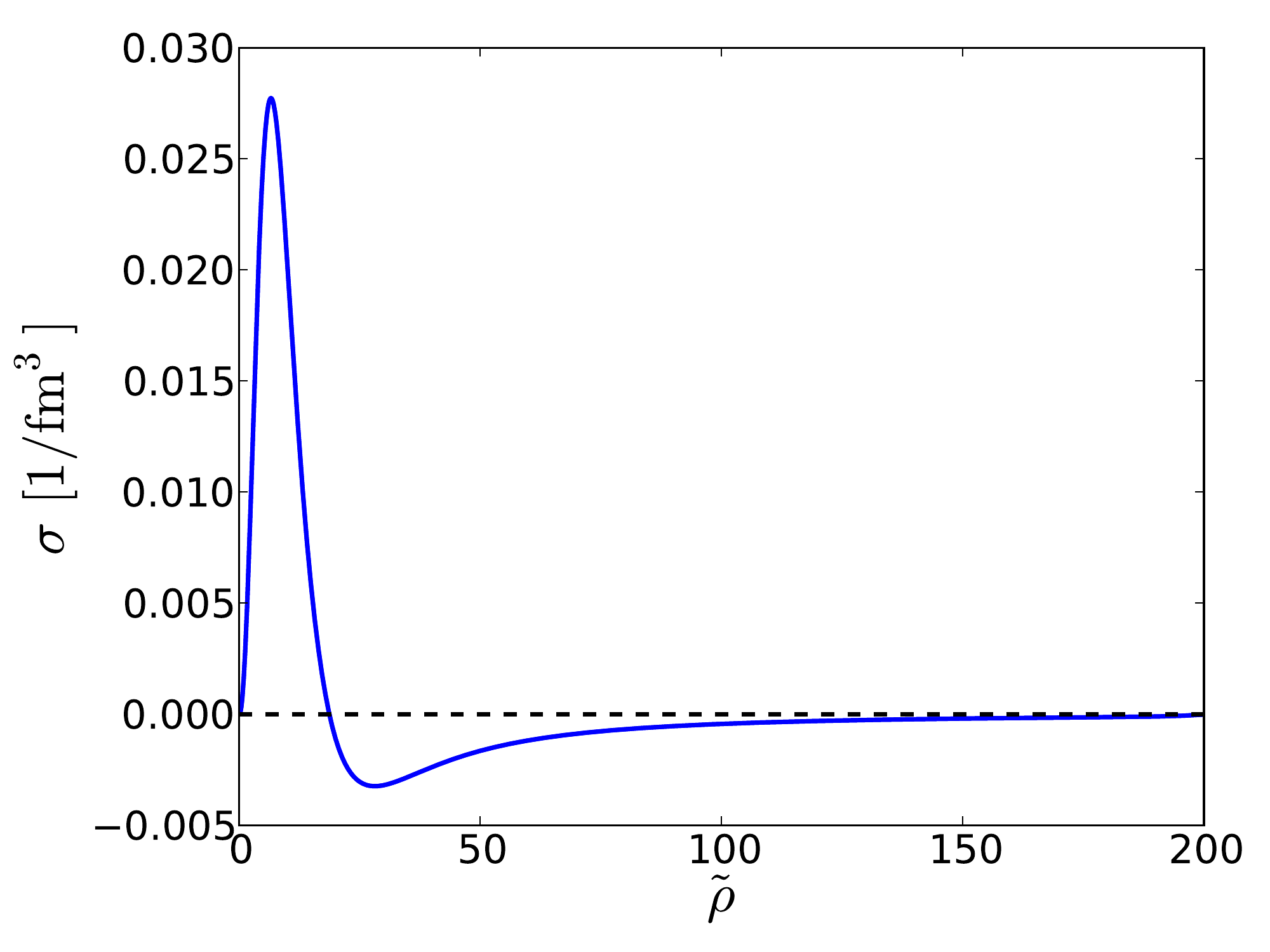}
\caption{Magnetization $\sigma(\tilde{\rho})$ for $\delta=0$ plotted as a function of the rescaled distance $\tilde{\rho}$ from the vortex centre. }
\label{Case1_mag_f1f2_F24}
\end{figure}

Here, 
$C_m = \frac{4}{9}N'(0)k_F^2$ and $N'(0) = m_n^2 /\left( 2\pi^2 k_F\right)$ is the number density of states $N =  m_n  k  /\left(2\pi^2\right)$ differentiated 
with respect to the energy $E=k^2/2m_n$ and evaluated at the Fermi surface $k=k_F$. For the numerical parameter values stated in Appendix~\ref{sec:GL}, $N'(0)$ takes the value  $1.341 \times 10^{-5} \text{MeV}^{-2} \text{fm}^{-2}$  and $C_m$ is computed to be $ 2.882 \times 10^{-5} \text{MeV}^{-2} \text{fm}^{-5}$. 
We display in Fig.~\ref{Case1_mag_f1f2_F24} the resulting magnetization $\sigma$~(\ref{Mag_sig}) for vanishing $\delta$ as a function of the rescaled distance $\tilde{\rho}$ from the vortex centre. The magnetization densities are evaluated using the vortex profile functions $f_1$, $f_2$ and $g$ plotted in Fig.~\ref{Case1_f1f2_F24}.
Substituting the vortex profile functions computed for vanishing $\delta$ in (\ref{magnetization}), we can also find an estimate for the magnetization of the system as a function of $\delta$. Integrating the magnetization (\ref{magnetization}) over the $xy$-plane gives the magnetic moment per unit  length
\begin{eqnarray}
\bm{m}(z)& =& \frac{\gamma_n \hbar}{2} 2\pi \xi^2\int \tilde{\rho} d\tilde{\rho} \bm{ \sigma}(\tilde{\rho})\nonumber \\
&=&C_m\frac{|\alpha|}{6\beta} \frac{\gamma_n \hbar}{2} 2\pi \xi^2\int \tilde{\rho} d\tilde{\rho}  g(\tilde{\rho})(f_1(\tilde{\rho})-f_2(\tilde{\rho})){\rm cos}\delta\,\, \hat{\bm{z}}\,.\nonumber \\
\label{Mag_mom_len}
\end{eqnarray}
The integration in (\ref{Mag_mom_len}) can be performed numerically using the vortex profile function depicted in Fig.~\ref{Case1_f1f2_F24}.\\ The integral is found to take the following value 
$2\pi\int \tilde{\rho} d\tilde{\rho}  g(\tilde{\rho})(f_1(\tilde{\rho})-f_2(\tilde{\rho})) = -404.83$. 
Hence, for vanishing $\delta$ the magnetic moment per unit length is given by $m(z)\approx 4.196\times 10^{-19}\text{MeV}\text{fm}^{-1}\text{G}^{-1}\approx 0.6 \times 10^{-27} \text{J} \,\text{T}^{-1} \text{fm}^{-1}$. Therefore, the magnetic moment of a $10$ fm string is found to be comparable to the neutron magnetic moment.

\section{The long distance  Effective description of free energy of a vortex}\label{sec:EffAct}
In this section we discuss the effective description of collective excitations along a vortex line by using the so-called moduli approximation, originally used for monopoles \cite{Manton:1981mp} 
and applied to various topological solitons, e.~g.~\cite{Eto:2006uw}. At large distances, superfluid vortices can be characterized by Kelvin modes. 
When orientating the vortex axis along the $z$-axis, these two moduli correspond to translations in the $x$- and $y$- direction. Physically these modes can be interpreted as the massless Nambu-Goldstone excitations generated from the spontaneous breaking of the translational symmetry \cite{Kobayashi:2013gba,Takahashi:2015caa}. These modes are well studied in the literature of superfluids \cite{Donnelly:1991} and are not an issue of this paper.

In this article, we want to explore collective excitations localized around the vortex. We introduce a massive mode $\delta$ which may be important for understanding the spontaneous magnetization of the $^3 P_2$ vortex.   Interestingly, when we switch on the mode $\delta$ in Eq.~(\ref{magnetization}), the spontaneous magnetization changes inside the vortex core.  

 In this article we solved the equations of motion by imposing cylindrical symmetry, which makes the profile function independent of the $z$-coordinate. The parameter $\delta$ has been introduced as a constant phase in the off-diagonal element for a vortex solution in the $xy$-plane. As we know, the vortex profile functions eventually reach their ground state values (\ref{BCs}) at large distances from the vortex core and $g(\rho)$ forms a concentric ring around the vortex. So $g(\rho)$ becomes practically zero outside the vortex, see Fig.~\ref{Case1_f1f2_F24}. Therefore, we may treat $\delta$ as a collective coordinate (or a modulus) of a vortex and may take it as a function of $z$ in the effective free energy functional. In principle, $\delta$ also depends on time but here we are not considering time dependence.  In our case, $\delta$ is a slowly varying function of the $z$-coordinate.

The parameter $\delta$ depends on $z$ and the $z$ derivative of $\delta$ also contributes to the total effective energy.
We insert the ansatz
 \begin{align}
&& A^{(\rho, \theta, z)}_{\d} =
 e^{i\t} \left(
    \begin{array}{ccc}
      f_1^0(\rho) & ig_0(\rho)e^{i\delta(z)} & 0 \\
      ig_0(\rho)e^{ i\delta(z)} & f^0_2(\rho) & 0 \\
      0 & 0 & -f_1^0-f_2^0
    \end{array}
  \right)    \label{eq:ansatz}
\end{align}
into the free energy functional (\ref{freeenergydensity}) with the sixth order and magnetic term set to be zero and with the gradient and up to fourth order term given in Eqs.~(\ref{free_contri_grad}) and (\ref{4th_contri_pot}), respectively. With the ansatz (\ref{eq:ansatz}) substituted into (\ref{freeenergydensity}), the effective free energy takes the following form 
\begin{widetext}
 \begin{align}
 F= &\frac{4\alpha^2 \xi^4}{\beta G} \int \text{d}z\,\text{d}^2\r\, \left[\frac{1}{\xi^2}\left\{ \left(\frac{2g_0}{\r^2}\left(u_0 - 4v_0 \right) + \frac{1}{\r} (u_0 {g_0}' - {g_0}{u_0}')\right)(\cos\delta-1) 
 +\frac{g^2}{144} (u_0^2 + v_0^2)(\cos2\delta - 1)\right\} +   (\p_z\d)^2 g_0^2\right]\,,\label{effectiveaction1}
\end{align} 
\end{widetext}
where $f_1^0$, $f_2^0$ and $g_0$ are the vortex profile functions for zero $\delta $, see Fig.~\ref{Case1_f1f2_F24}. $u_0(\r)$ and $v_0(\r)$  are defined as 
 $u_0 =  f_1^0 + f_2^0$ and $v_0 =  f_1^0 - f_2^0$, respectively. The rescaled radial distance $\r$ is given by $\r = \rho/\xi$, where $\xi$ is the coherence length of the system. Eq.~(\ref{effectiveaction1}) can be expressed more neatly as,
 \begin{align}
 \label{Feff}
&F_{\text{eff}} = \int \text{d}z\,\,\frac{4\alpha^2 \xi^4}{\beta G} \left[\frac{1}{\xi^2}\left\{ \lambda_1(\cos\delta-1) +\lambda_2(\cos2\delta - 1)\right\} \R.\nonumber\\
&\L.\qquad\qquad\phantom{xxxxxxxxxxxxxxxxxxxxxxxxxxxxxxxx}+   (\p_z\d)^2\lambda_3\large\right]\,,
\end{align}
where the coefficients are defined as 
\begin{subequations}
\begin{align}
\lambda_1& = \int d^2 \r \left(\frac{2g_0}{\r^2}\left(u_0 - 4v_0 \right) + \frac{1}{\r} (u_0 g_0' - g_0u_0')\right)\,, \,\,\\
\lambda_2 &= \int d^2 \r\,\,\frac{g^2}{144} (u_0^2 + v_0^2)\,,\\
\lambda_3 &= \int d^2 \r g_0^2\,.
\end{align}
\end{subequations}
 The coefficients $\lambda_1, \lambda_2$ and  $\lambda_3$ have been evaluated numerically to be given by 
 \begin{align}\label{lambdaval}
\lambda_1 = -4.889\,, \quad \lambda_2 =  0.411\,,\quad \lambda_3  = 17.748\,.
\end{align}

\begin{figure}[!htb]
\includegraphics[totalheight=7.1cm]{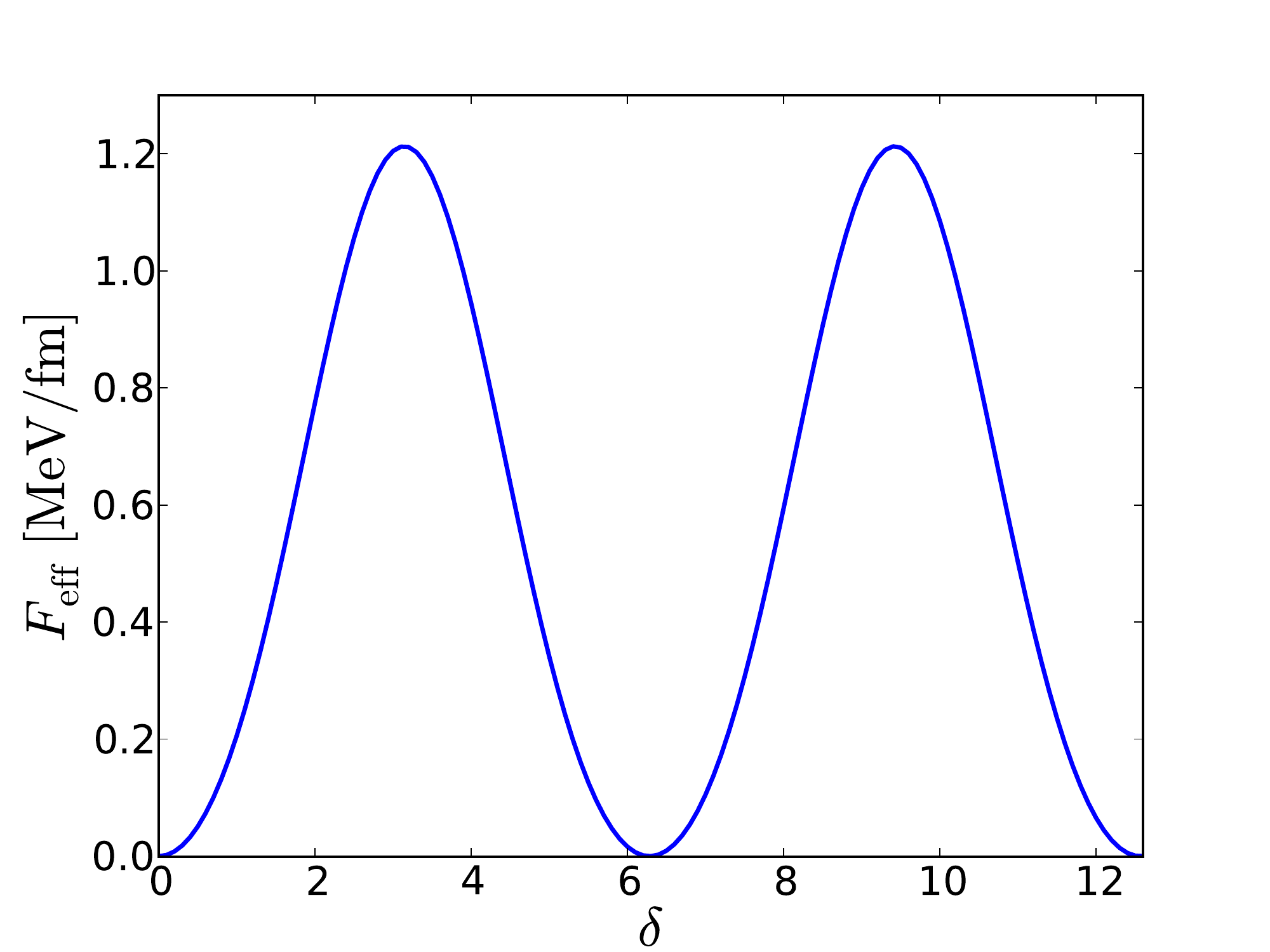}
\caption{We display the effective free energy potential $F_{\text{eff}}$ (\ref{Feff}) as a function of the $\delta$ parameter. We evaluate the free energy for the $\lambda$ coefficients listed in Eq.~(\ref{lambdaval}). The GL parameters are given by $\alpha=-0.1$ and $\beta = 2.57664\, \text{(MeV\, fm)}^{-2}$. The coherence length of the system is set to $30.3634\,\text{fm}$. Note that the energy values are given in units of MeV $\text{fm}^{-1}$.}
\label{Effective_plot}
\end{figure}
Note that the signs and numerical values of the $\lambda$ coefficients are very important in this work. The graph of the effective potential (\ref{Feff}) should exhibit a bump at the bottom because of the negative sign of the $\lambda_2$ contribution. However, as can be seen from Fig.~\ref{Effective_plot}, the minima occur at $\d = 2\pi n$ with $n$ being an integer. To get any local extrema other than $n\pi$, $\delta$ has to satisfy the equation 
\begin{eqnarray}
\cos\d = \frac{|\lambda_1|}{4\lambda_2}\,.
\end{eqnarray}
It can be checked that the numerically evaluated $\lambda$ values (\ref{lambdaval}) yield $\frac{|\lambda_1|}{4\lambda_2} > 1$. Hence, the obtained $\lambda$ coefficients do not support any other local minima or maxima. This finding supports our expansion of the system around $\d =0$.

\section{Double sine-Gordon Kink solution on a vortex}\label{sec:DGkink}
In the last section, we have derived the effective free energy functional (\ref{Feff}) that describes the low-energy dynamics of the system along with other dynamical terms.  Since in this article we are solely interested in time independent configurations, we discuss in the following the static solutions of the system. Note that the system (\ref{Feff}) is known as the double sine-Gordon model. However, the potential graph shown in Fig.~\ref{Effective_plot} almost agrees with the sine-Gordon case because of $|\lambda_2|<<|\lambda_1|$, see Eq.~(\ref{lambdaval}). Here, we shall derive the kink solution  and describe the physics behind the existence of the kink.  
We use the technique similar to Bogomol'nyi-Prasad-Sommerfield (BPS) monopoles \cite{Bogomolny:1975de, Prasad:1975kr}.

     Let us  rewrite the free energy (\ref{Feff}) as,
\begin{align}
 \label{Feff21}
&F_{\text{eff}} = \frac{4\alpha^2 |\lambda_1| \xi^2}{\beta G} \int \text{d}z\,\, \left[\left\{ (1 - \cos\delta) + \frac{\lambda}{2} (\cos2\delta - 1)\right\} \R.\nonumber\\&\L.+  \frac{\xi^2\lambda_3}{|\lambda_1|} (\p_z\d)^2\right]\,, 
\end{align}
where
$ \lambda = \frac{2\lambda_2}{|\lambda_1|}\,.$
The numerical value of $\lambda$ is evaluated to be $0.168735$. After rescaling $z$, 
the effective free energy in Eq.~(\ref{Feff21}) takes the form 
\begin{align}
 \label{Feff2}
&F_{\text{eff}} =& \frac{M_{\rm E}}{2} \int \text{d}\zeta\,\, \left[\left\{ 2(1 - \cos\delta) + \lambda (\cos2\delta - 1)\right\} +  (\p_{\zeta}\d)^2\right]\, , 
\end{align}
where 
 $\zeta = \sqrt{ \frac{|\lambda_1|}{2 \lambda_3}} \,\frac{z}{\xi}\, $
and
\begin{align}
M_E = \frac{4\alpha^2\xi^3\sqrt{2|\lambda_1| \lambda_3}}{\beta G}\,.
\end{align}

Although the second term in Eq.~(\ref{Feff2}) is not positive, the potential part of Eq.~(\ref{Feff2}) is positive definite.  So, we can use the BPS technique to find the kink solution. To do that we write the $F_{\text{eff}}$ as,
\begin{align}
 \label{Feff3}
F_{\text{eff}} =& \frac{M_E}{2} \int \text{d}\zeta \left[ \left\{\p_{\zeta}\d \pm\sqrt{2(1 - \cos\delta) + \lambda (\cos2\delta - 1)}\right\}^2 \right. \nonumber \\& \left.+ M_E\, \left|  \p_\zeta\d \sqrt{  2(1 - \cos\delta) + \lambda (\cos2\delta - 1)}\right|.   \right]\, 
\end{align}
Then $F_{\text{eff}}$ satisfy the inequality
\begin{align}
F_{\text{eff}} \ge E_{\rm{BPS}} =& M_E \left| \int d\zeta\p_\zeta\d \sqrt{ 2(1 - \cos\delta) + \lambda (\cos2\delta - 1)}\right|.
\end{align}
The minimum energy kink configurations would saturate the bound and  satisfy the so-called
 BPS equation
\begin{eqnarray}
\label{bpseqn}
\p_{\zeta}\d = \pm \sqrt{ 2(1 - \cos\delta) + \lambda (\cos2\delta - 1)}.
\end{eqnarray}
 It is easy to show that the solutions of above equation also satisfy the original equations of motion. The (anti)soliton solution of the  BPS equation can be written as,
\begin{eqnarray}\label{DSG_soliton}
\d(\zeta)_{}\pm &=  \pi \pm 2 \tan^{-1} \left(\Lambda\,\sinh \Lambda \zeta\right)\, ,
\end{eqnarray}
where
$
\Lambda = \sqrt{1 - 2\lambda}\,
$
is a real positive constant.
\begin{figure}[!htb]
\includegraphics[totalheight=7.1cm]{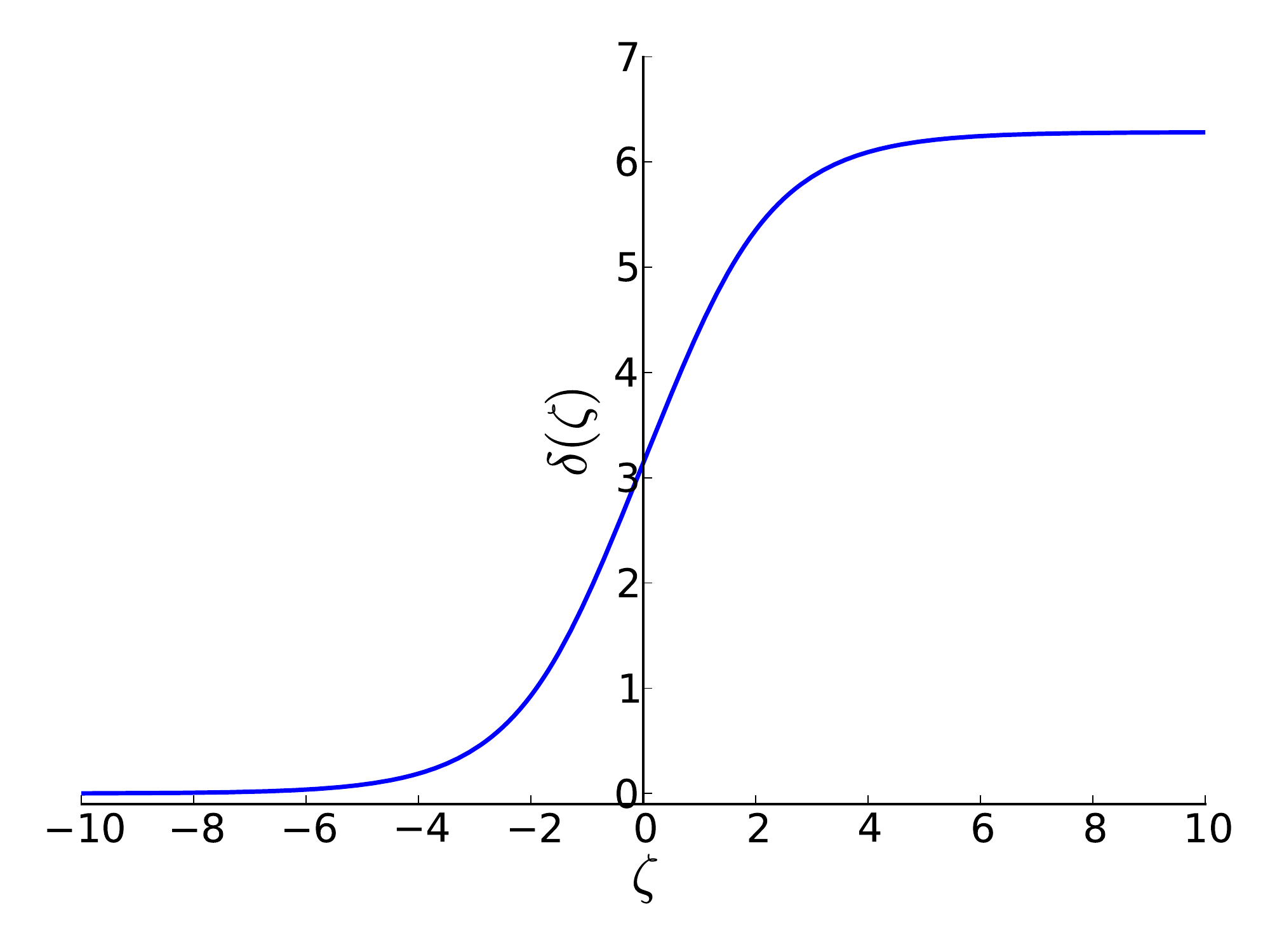}
\caption{The double sine-Gordon kink solution $\delta(\zeta)$ (\ref{DSG_soliton}) shown for $\Lambda = 0.813959$.}
\label{DSG_solution}
\end{figure}
It can be easily seen that for $\lambda = 0$, the above solution reduces to the usual sine-Gordon kink
$\delta_{sG}(\zeta) = 4\arctan e^{\pm\zeta}$.

The BPS energy bound for the (anti)soliton  can be expressed as
\begin{align}
E_{\rm{BPS}} =& M_E \int^{\d(\infty)=2\pi}_{\d(-\infty)=0} d\d \sqrt{ 2(1 - \cos\delta) + \lambda (\cos2\delta - 1)}\nonumber\\
=&  \frac{M_E}{\sqrt{2\lambda}} \int^{\sqrt{2\lambda}}_{-\sqrt{2\lambda}} dt \sqrt{ 1 -  t^2} \nonumber \\ 
 =& 1.88 \, M_E\, .
\end{align}
For a nuclear matter density of $\rho_n=6\times 10^{17}\, \text{kg}\,\text{m}^{-3}$ \cite{Richardson:1972xn} and a temperature of $T=0.16$ MeV (as assumed in Appendix~\ref{sec:GL} and Table~\ref{table-coeffi}), the total kink energy can be evaluated as 
\begin{eqnarray}
E_{\rm{BPS}} = 93.3\,\, \text{MeV}\,.
\end{eqnarray}
We display in Fig.~\ref{DSG_solution} the double sine-Gordon kink solution $\delta(\zeta)$ obtained for the $\lambda$ values given in Eq.~(\ref{lambdaval}).
\begin{figure}[!htb]
\includegraphics[totalheight=7.1cm]{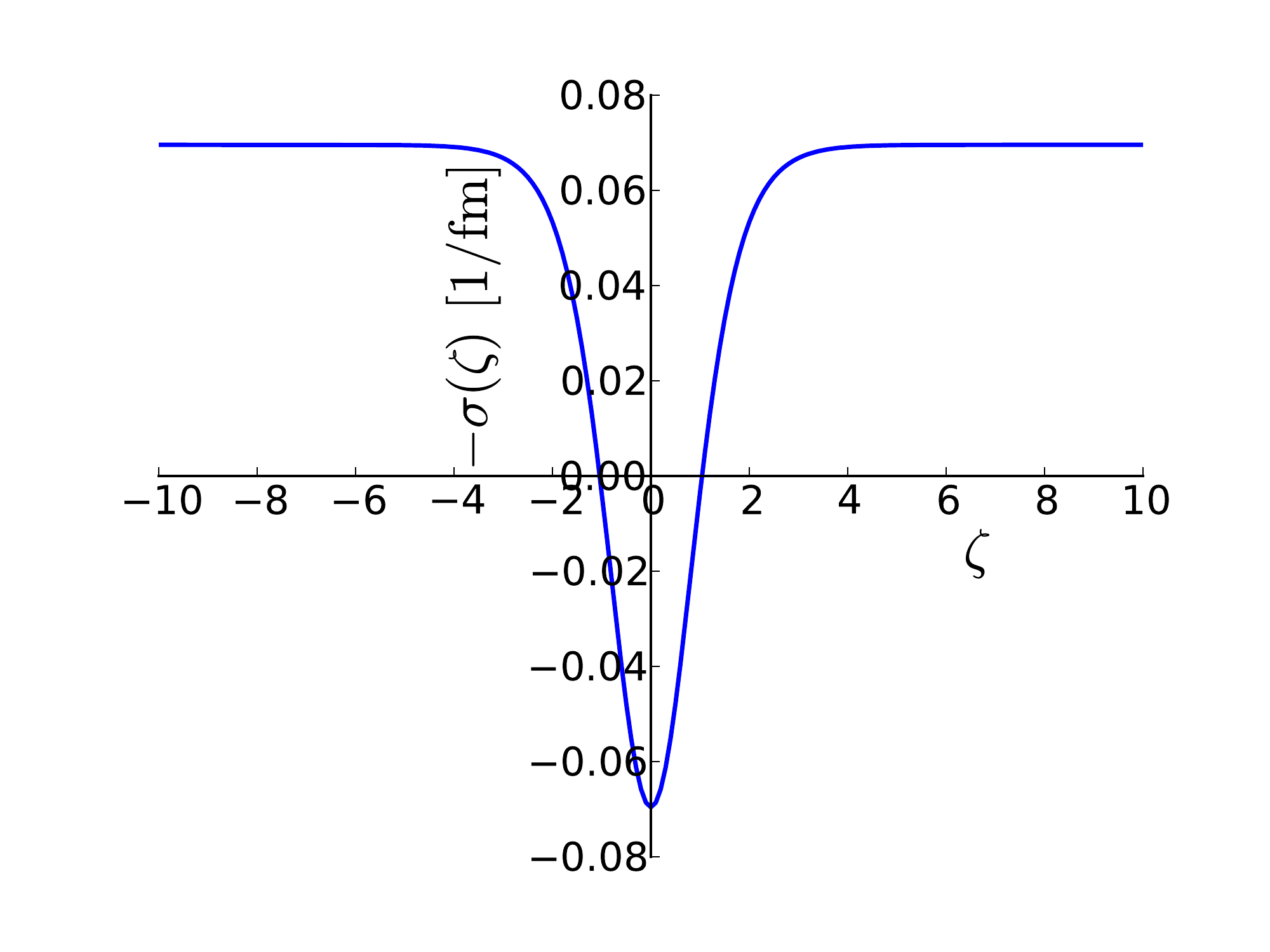}
\caption{The magnetization $\left(\frac{2m(\zeta)}{|\gamma_n|\hbar}= -\sigma(\zeta)\right)$ plotted as a function of the rescaled $z$-coordinate $\zeta$. }
\label{Sigma_Z}
\end{figure}
In Fig.~\ref{Sigma_Z}, we plot the change in the vortex core magnetization in the presence of the kink solution along the $z$-axis. When substituting (\ref{DSG_soliton}) in Eq.~(\ref{Mag_mom_len}), it is observed that the magnetization changes its value continuously and reaches its opposite value at the center of the kink.

\begin{figure}[htb]
\includegraphics[ height= 7.1 cm]{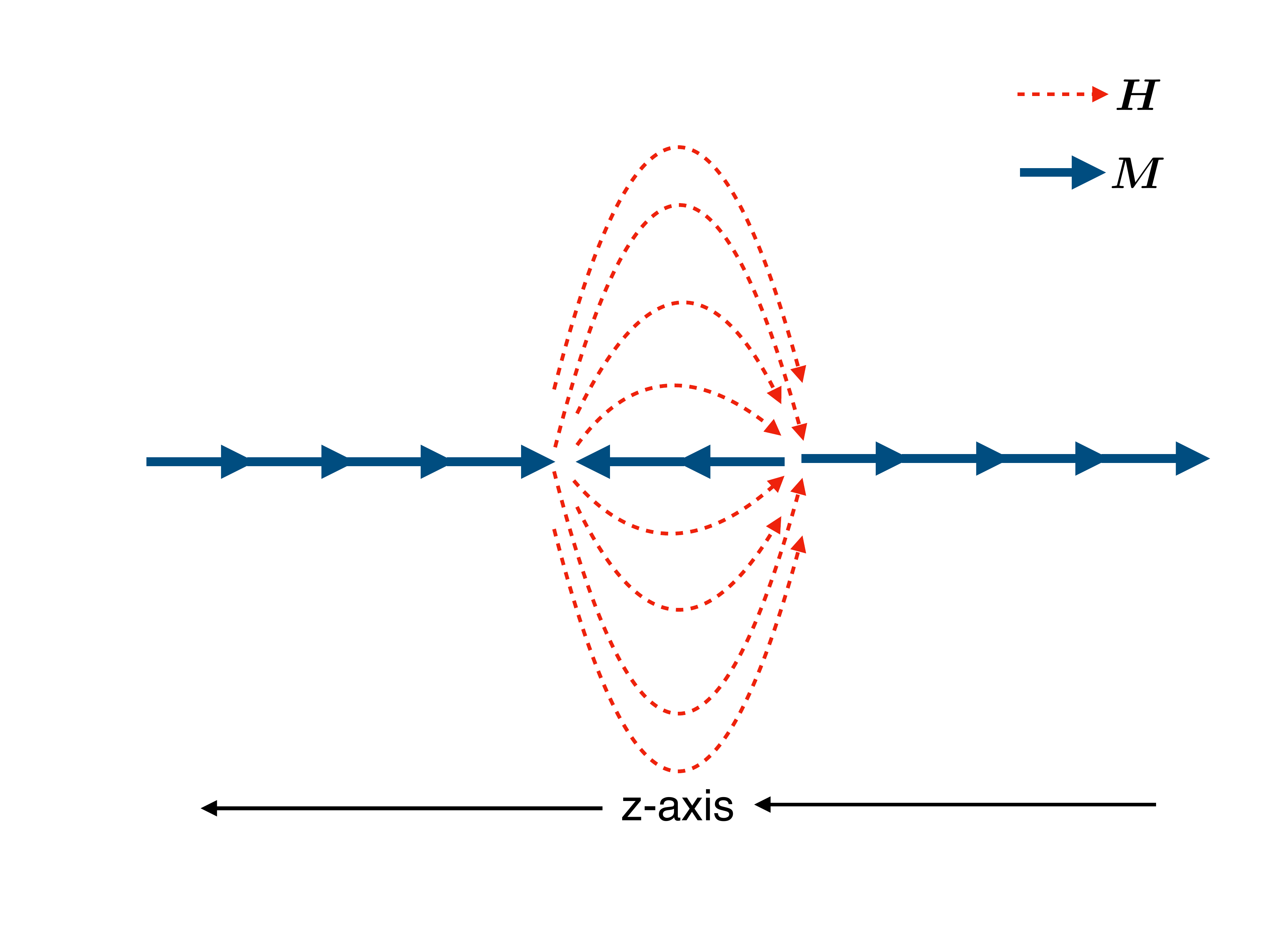}
\centering
\caption{A schematic figure  of a magnetic field configuration to illustrate the flipped magnetic moment and the leakage. The dotted red arrows and  the solid blue arrows denote  the  magnetic field $\bm H$ and  magnetic moment $\bm M$, respectively.}
\label{Dm}
\end{figure}

Note that in this article we observed a  sign flip of the trapped magnetic moment inside the core of a $^3P_2$ neutron superfluid vortex. For this case it is natural to think of the system as a chain of aligned magnetic moments with a flipped moment at the position of the kink. Hence, $\nabla\cdot \bm M \ne 0$ on the junctions, where $\bm M$ is defined in Eq.~(\ref{magnetization}).  Since the total magnetic field $\bm B $ is divergenceless, $\nabla\cdot \bm B = 0$, there would be a leakage of a magnetic field, say $\bm H$, from the junctions. In this case   $\nabla \cdot\bm H \ne 0$  at the junctions and the total magnetic field is defined as $\bm B = \left(\bm H +  4 \pi \bm M\right)$ .
We schematically  show in Fig.~\ref{Dm} an example of a magnetic field configuration with a flipped magnetic moment together with the resulting leakage of the magnetic field. 

\section{Conclusions}\label{sec:Con}

In this article, we have derived an effective theory of $^3P_2$ neutron superfluid vortices at large distances and expressed solely in terms of the phase $\delta$ of the vortex profile function $g$. $g$ lives in one of the off-diagonal components of the tensorial $^3P_2$ order parameter $A_{\alpha\,i}$. Here, $\delta$ plays a crucial role as it gives a nonzero spontaneous magnetisation in the core region of $^3P_2$ neutron superfluid vortices for the angular independent profile function $g$. Variations in $\delta$  change the induced magnetic moment accordingly. In our analysis we have found that at low energies the $^3P_2$ superfluid vortices are described by a double sine-Gordon model in addition to the well-known Kelvin (translational zero) modes part. We have derived a kink solution of this particular double sine-Gordon model and have determined its BPS energy bound. We have found that the vortex core magnetisation flips its direction at the kink on the vortex.

Since the number of vortices in a rotating NS is of the order of 
$10^{19}$ and the NS radius is about 10 km, 
there should exist a large number of magnetic strings in the NS interior like the ones discussed in this paper. 
A pair of kink and anti-kink may be created spontaneously during the creation of the vortex through phase transition.
Possible implications of these objects for NS physics should be one of 
the most important future problems.

In this article, we have neglected higher order terms (esp. sixth order terms) in the GL free energy. We also did not take into account the effect of an external magnetic field. However, in the core of neutron stars where it is most likely to find neutron superfluid, the inclusion of high external magnetic fields and of the sixth order term would be important. So it would be interesting to study the effective free energy in the presence of high external magnetic fields and with nonzero sixth order term included. We have also considered only static configuration in this article.
In order to discuss dynamics, we have to include time dependence.
To this end, we need a time dependent GL equation, 
which has not been obtained yet. For the time-dependent problem, the phase $\delta$ has to be promoted to a time-dependent function and can be treated as a field which lies on the two-dimensional string world sheet. Fluctuations of $\delta$ may indicate changes in the magnetization on the world sheet. In the case of conventional superfluids, 
Kelvin modes will propagate with quadratic dispersion relation 
for small system sizes, see e.~g.~Ref.~\cite{Kobayashi:2013gba}.
However, it is well known that for  large system sizes the dispersion relation is given by 
$\epsilon = k \log k$ with wave vector $k$.
Recently, a formula for arbitrary system sizes 
has been obtained in Ref.~\cite{Takahashi:2015caa}.
On the other hand, the gapfull mode $\delta$ was previously unknown.
We expect $\delta$ to have relativistic dynamics.

In this article, we have discussed only the integer vortex. 
For strong magnetic fields, the ground state is in the $D_4$-biaxial nematic phase 
admitting half-quantized non-Abelian vortices 
\cite{Masuda:2016vak}. 
The vortex core magnetization is ten times larger than that of the integer vortex. 
For this case we should also derive the low-energy collective coordinate approximation 
and construct a magnetic lump.

\section*{Acknowledgements}

This work is supported by the Ministry of Education, Culture, Sports, Science (MEXT)-Supported Program for the Strategic Research Foundation at Private Universities ``Topological Science'' (Grant No. S1511006). C.~C. acknowledges support as an International Research Fellow of the Japan Society for the Promotion of Science (JSPS). Some of the work of M.~H. was undertaken at the Department of Mathematics and Statistics, University of Massachusetts, financially supported by FP7, Marie Curie Actions, People, International Research Staff Exchange Scheme (IRSES-606096).
The work of M.~N. is supported in part by 
JSPS Grant-in-Aid for Scientific Research (KAKENHI Grant No. 16H03984), 
and by a Grant-in-Aid for
Scientific Research on Innovative Areas ``Topological Materials
Science'' (KAKENHI Grant No.~15H05855) and ``Nuclear Matter in Neutron
Stars Investigated by Experiments and Astronomical Observations''
(KAKENHI Grant No.~15H00841) from the the Ministry of Education,
Culture, Sports, Science (MEXT) of Japan.


\appendix 
\section{Ginzburg-Landau Description for $^3P_2$ Superfluids}\label{sec:GL}

In this appendix, we briefly discuss the Ginzburg-Landau (GL) construction of $^3P_2$ superfluids and the associated GL parameter values.
The  Hamiltonian with $^3P_2$ contact interaction can be written as
\begin{align}
H &= \int d^3 \rho\ \psi^{\dagger}\left(-\frac{\bm{\nabla}^2}{2M}-\mu\right)\psi
-\frac{1}{2}gT^{\dagger}_{\alpha \beta}(\bm{\rho})T_{\alpha \beta}(\bm{\rho}),  \\ 
T^{\dagger}_{\alpha \beta} (\bm{\rho}) 
 &= \psi^{\dagger}_{\sigma}(\bm{\rho})
    (t^{\ast}_{\alpha \beta})_{\sigma \sigma'}(\bm{\nabla})
    \psi^{\dagger}_{\sigma'}(\bm{\rho}) ,
\label{hamiltonian}
\end{align}  
where $\bm{\rho}$ denotes the space coordinates, $\psi$ is a neutron field, $\mu$ is a baryon chemical potential, 
$M$ is the neutron mass, and $g(>0)$ is the coupling constant. 
Here, $\alpha,\beta$ are the space indices, and 
the tensor $T_{\alpha\beta}$ represents  the $^3P_2$ pair creation and  annihilation operator. 
The differential operator $t$ is defined as
\begin{align}
(t_{\alpha \beta})_{\sigma\sigma'}(\bm{\nabla})
&=&\frac{1}{2}((S_{\alpha})_{\sigma\sigma'}\nabla_{\beta}
+\nabla_{\alpha}(S_{\beta})_{\sigma\sigma'}) \nonumber \\&&
-\frac{1}{3}\delta_{\alpha \beta}(\bm{S})_{\sigma\sigma'}\cdot \bm{\nabla},
\end{align}
and $\bm{S}$ is given by $(S_{\alpha})_{\sigma\sigma'} = 
i(\sigma_y \sigma_{\alpha})_{\sigma\sigma'}$ with $\alpha=x,y,z$ and pauli matrices $\sigma$.

The $^3P_2$ order parameter can be represented in general  as a  $3 \times 3$ traceless symmetric  complex tensor $A_{\mu i}$, which is defined in Refs.~\cite{Fujita01011973, Sauls:1982ie} as
\begin{eqnarray}
\Delta=\sum_{\alpha i} i\sigma_{\alpha}\sigma_y A_{\alpha i} k_i,
\end{eqnarray}
where $\Delta$ is the gap parameter. 
As above, Greek subscripts stand for spin indices while Roman indices denote the spatial coordinates. The tensor $A_{\alpha i}$ is related to $T_{\alpha\beta}$ as given in Ref.~\cite{Richardson:1972xn},
\begin{eqnarray}
A_{\alpha\beta} = g \langle T_{\alpha\beta}(\bm{\rho})\rangle_T,
\end{eqnarray}
where the angular brackets are defined as  an ensemble average in the grand canonical ensemble.

\begin{table*}[t!]
\begingroup
\renewcommand{\arraystretch}{3}
 \begin{tabular}{|c|c|c|} \hline\hline
    $\alpha$ & $K_1=K_2$ & $\beta$ 
     \\ \hline 
    $\displaystyle{\frac{T-T_c}{2T_c}}$ & 
    $\displaystyle{\frac{7\zeta (3)}{240M_n^2}\frac{N(0)G}{(\pi T_c)^2}k_F^4}$ & 
    $\displaystyle{\frac{7\zeta (3)}{60}\frac{N(0)G}{(\pi T_c)^2}k_F^4}$ 
    \\ \hline
    $-0.1$& $1106.32 \,\,\text{fm}^2$ & 2.57664 $(\text{MeV}\,\text{ fm})^{-2}$ 
    \\\hline\hline
  \end{tabular}
\endgroup
\caption{Summary of the GL parameters computed in the weak coupling limit. We refer the reader to the last paragraph of section~\ref{sec:GL} for the numerical values of all model parameters used in our numerical simulations.}
\label{table-coeffi}
\end{table*}

\section{The Large Distance Bahaviour of the Order Parameter for a Vortex}{\label{vortex:order}
 An ansatz for the order parameter $A$ of a vortex state has been developed first in Refs.~\cite{Fujita01091972,Richardson:1972xn}. Here, we briefly sketch its derivation. For a detailed derivation we refer the interested reader to Ref.~\cite{saulsthesis:1980ab}. For an isolated vortex solution with cylindrical symmetry and  phase $ m \theta$, where $m$ measures the circulation about the vortex line, we first choose the order parameter at  large distances  by minimising the potential  in Eq.~(\ref{4th_contri_pot}) as
\begin{eqnarray}
 A &
\sim
& C(\eta)e^{im\theta}
 R(\theta) \left(
    \begin{array}{ccc}
      \eta & 0 & 0 \\
      0 & -  (1 + \eta) & 0 \\
      0 & 0 &  1
    \end{array}
  \right)R^T(\theta)\,, \\
  C(\eta) &=&  \sqrt{\frac{|\alpha|}{\beta(1 + \eta^2 + (1 + \eta)^2)}} \,.  \label{eq:ansatz_1_app}\nonumber
\end{eqnarray}
Here $R(\theta)$ is the rotation matrix which rotates the axis from $(\rho, \theta, z)$ into $(x, y, z)$. Unlike in the case of the ground state,  the order parameter can also have extra contributions from the kinetic terms in the free energy functional. 
Both the $K_1$ and $K_2$ terms in Eq.~(\ref{free_contri_grad})
give logarithmic contributions to the total energy. Minimizing the coefficient of the logarithmic term 
breaks the ground state degeneracy for all $-1 \le \eta \le -\half$
and the lowest energy is achieved for $\eta = 6 -\sqrt{43} \sim -0.557$. Hence, for a vortex configuration of minimal energy we may choose the matrix order parameter of a vortex state as given in Refs.~\cite{Fujita01091972,Richardson:1972xn,Sauls:1982ie,saulsthesis:1980ab}
 \begin{eqnarray}
 A 
\sim
\sqrt{\frac{|\alpha|}{6\beta}}e^{i\theta}
 R(\theta) \left(
    \begin{array}{ccc}
      f_1 & 0 & 0 \\
      0 & f_2 & 0 \\
      0 & 0 & -f_1-f_2
    \end{array}
  \right)R^T(\theta)\,  \,,  \label{eq:ansatz_1_app}
\end{eqnarray}
where the boundary conditions for the profile functions are chosen according to the large distance behaviour in Eq.~(\ref{eq:ansatz_1_app}). }

In Table~\ref{table-coeffi}, we summarise the coefficients $\alpha$, $\beta$, $K_1$ and $K_2$ (the GL parameters) calculated in the weak coupling limit by considering only the excitations around the Fermi surface \cite{Fujita01011973,Richardson:1972xn, saulsthesis:1980ab}. In this limit, $K_1$ and $K_2$ in Eq.~(\ref{free_contri_grad}) take the same value. To simplify comparison with the works \cite{Richardson:1972xn, Sauls:1982ie, Masuda:2015jka}, the GL parameters are calculated for a nuclear mass density of $\rho_n=6\times 10^{17}\, \text{kg}\,\text{m}^{-3}$ \cite{Richardson:1972xn}, Fermi momentum $k_F = 2.1987 \,\text{fm}^{-1}$, nucleon mass $\text{m}_n = 0.02413\, \text{MeV}^{-1} \text{fm}^{-2}$, a critical temperature for ${}^3P_2$ pairing of $T_C=0.2\,\text{MeV}$ and a temperature of $T=0.8T_C$.

\section{Equations of Motion}\label{App_Eom}

In this appendix, we list explicitly all the vortex equations together with the imposed boundary conditions. Note that all vortex equations are given in the cylindrical basis ($n=1$).

The equations of motion derived from (\ref{Free_case_1}) are given by 
\begin{subequations}\label{EOM_n1m0}
\begin{align}
&\r^2 \left(4 u^{\prime\prime}+v^{\prime\prime}\right)+\r (4 u^\prime+3v ^\prime-4g^\prime)-\left(4u-v+4g\right)\nonumber\\&-\frac{\r^2}{144}\left(\frac{36 \alpha }{|\alpha|}+9u^2+3 v^2+20g^2\right) u=0\,,\\
&\r^2 \left(u^{\prime\prime}+2v^{\prime\prime}\right)+\r \left(2v^\prime-u^\prime\right)- \left(10v-u-16g\right)\nonumber\\&-\frac{\tilde{\rho}^2}{144}\left(\frac{12\alpha }{|\alpha|}+ 3u^2+v^2+
12 g^2\right)v=0\,,\\
&4\r^2 g^{\prime\prime}+2\r (2g^\prime + u^\prime)+2\left(4v-u-10 g\right)  \nonumber\\&-\frac{\r^2}{72}\left(\frac{12\alpha}{|\alpha|} + 5u^2+3v^2+4g^2\right)g=0\,.
\end{align}
\end{subequations}

All profile function $u$, $v$ and $g$ are required to vanish at the origin. At spatial infinity we impose the following boundary conditions which are compatible to the above equations of motion
\begin{subequations}\label{BC_n1m0}
\begin{eqnarray}
&&u^2 = (f_1+f_2)^2 = \frac{6}{(\eta^2 + (1+\eta)^2+1)},\\
&&v^2=(f_1-f_2)^2 = \frac{6(2\eta+1)^2}{(\eta^2 + (1+\eta)^2+1)},\\
&&u'=v'=g'=g=0.
\end{eqnarray}
\end{subequations}
Hence, with the vortex boundary state selected to be given by the lowest energy state $\eta=-0.557$ (due to the fourth order term), the linear combinations $u=f_1 +f_2$ and $v=f_1 - f_2$  have to fullfill the boundary conditions $u(\infty)=1.9956$ and $v(\infty)=0.2275$.


\end{document}